\documentclass[aps,reprint,superscriptaddress,a4paper,showpacs,showkeys,prl]{revtex4-1}
\usepackage{graphicx} 
\usepackage{dcolumn} 
\usepackage{bm}
\usepackage{hyperref} 
\usepackage{color} 
\usepackage[dvipsnames]{xcolor}
\usepackage{amsmath}
\usepackage{amssymb} 
\usepackage[normalem]{ulem}

\begin{document}

\title{Deviation of viscous drops at chemical steps} \author{Ciro
  Semprebon} \email{ciro.semprebon@ed.ac.uk} \affiliation{Department
  of Mechanical Engineering, University of Edinburgh, Edinburgh EH9
  3FB, UK}

\author{Silvia Varagnolo} \affiliation{Dipartimento di
  Fisica e Astronomia `G.Galilei' - DFA, Universit\`a di Padova, via Marzolo
  8, 35131 Padova, Italy} 

\author{Daniele Filippi} \affiliation{Dipartimento di
  Fisica e Astronomia `G.Galilei' - DFA, Universit\`a di Padova, via Marzolo
  8, 35131 Padova, Italy} 

\author{Luca Perlini} \affiliation{Dipartimento di
  Fisica e Astronomia `G.Galilei' - DFA, Universit\`a di Padova, via Marzolo
  8, 35131 Padova, Italy}

\author{Matteo Pierno} \affiliation{Dipartimento di
  Fisica e Astronomia `G.Galilei' - DFA, Universit\`a di Padova, via Marzolo
  8, 35131 Padova, Italy}

\author{Martin Brinkmann} \affiliation{Experimental Physics, Saarland
  University, 66123 Saarbr\"ucken, Germany}

\author{Giampaolo Mistura} \email{giampaolo.mistura@unipd.it}
\affiliation{Dipartimento di
  Fisica e Astronomia `G.Galilei' - DFA, Universit\`a di Padova, via Marzolo
  8, 35131 Padova, Italy}

\begin{abstract}
  We present systematic wetting experiments and numerical simulations
  of gravity driven liquid drops sliding on a plane substrate decorated with a
  linear chemical step. Surprisingly, the optimal direction to observe 
  crossing is not the one perpendicular to the step, but a finite angle that 
  depends on the material parameters. We computed the landscapes of
  the force acting on the drop by means of a contact line mobility model 
  showing that contact angle hysteresis dominates the dynamics at the step 
  and determines whether the drop passes onto the lower substrate. This analysis 
  is very well supported by the experimental dynamic phase diagram in terms 
  of pinning, crossing, sliding and sliding followed by pinning.
\end{abstract}

 \maketitle


Contact line motion on solid substrates is crucial in many natural
phenomena and technological processes  
which may have a wide variety of practical applications in daily life, industry and agriculture
like rain drops dragged on car windscreens or inkjet printing 
\cite{Bonn2009a,quere2008,liu2010}.  Macroscopic wetting defects such
as chemical patches \cite{Varagnolo2013,Varagnolo2014,nakajima2013}, topographic
structures \cite{Kalinin2009,baret2007} and embedded electrodes
\cite{tMannetje2011,tMannetje2014,brunet2010} have been proposed to control drop
motion.  Defects with anisotropic shape or spatial distribution may
also induce anisotropic \cite{Bliznyuk2010,Courbin2007} or even
unidirectional \cite{Chu2010,Malvadkar2010} liquid spreading.
Macroscale effects of microscopic defects, instead, are often
quantified by means of global descriptors such as the contact angle
hysteresis \cite{Semprebon2012}. Predicting the effect of
microscopic defects onto drop motion is a challenging task,
and the majority of previous studies are restricted to lower
dimensions \cite{Joanny1990,Savva2013,Herde2012}. Consequently, to
simplify the modelling of drop dynamics in the presence of macroscopic
defects \cite{Kusumaatmaja2007,Sbragaglia2013,Cavalli2015}, the
intrinsic contact angle hysteresis is often neglected, despite its 
ubiquity. For example in Refs.~\cite{Varagnolo2013,tMannetje2014},
the effect of the wetting heterogeneity on the drop dynamics is
modelled by a spatial variation of the surface energy. Such models are
reasonable only if the contact angle hysteresis interval is narrow
compared to the wetting contrast between the substrates.\\
\indent In this letter, we present experiments and numerical simulations 
of viscous drops moving on substrates with a large contact angle hysteresis. 
More specifically, we tracked the deflection of gravity driven drops at 
a straight wettability step \cite{Nilsson2012}, which can be regarded as 
the limiting case of a pattern of chemical stripes \cite{Suzuki2008,Varagnolo2013}. 
Our work demonstrates that, in the presence of macroscopic defects, 
the effects of contact line friction are not negligible and can lead to
counter-intuitive results: unexpectedly, the drop crosses more easily when the step 
is not perpendicular to the sliding direction.\\
\begin{figure}[htb]
\centering \includegraphics[width=0.84\columnwidth]{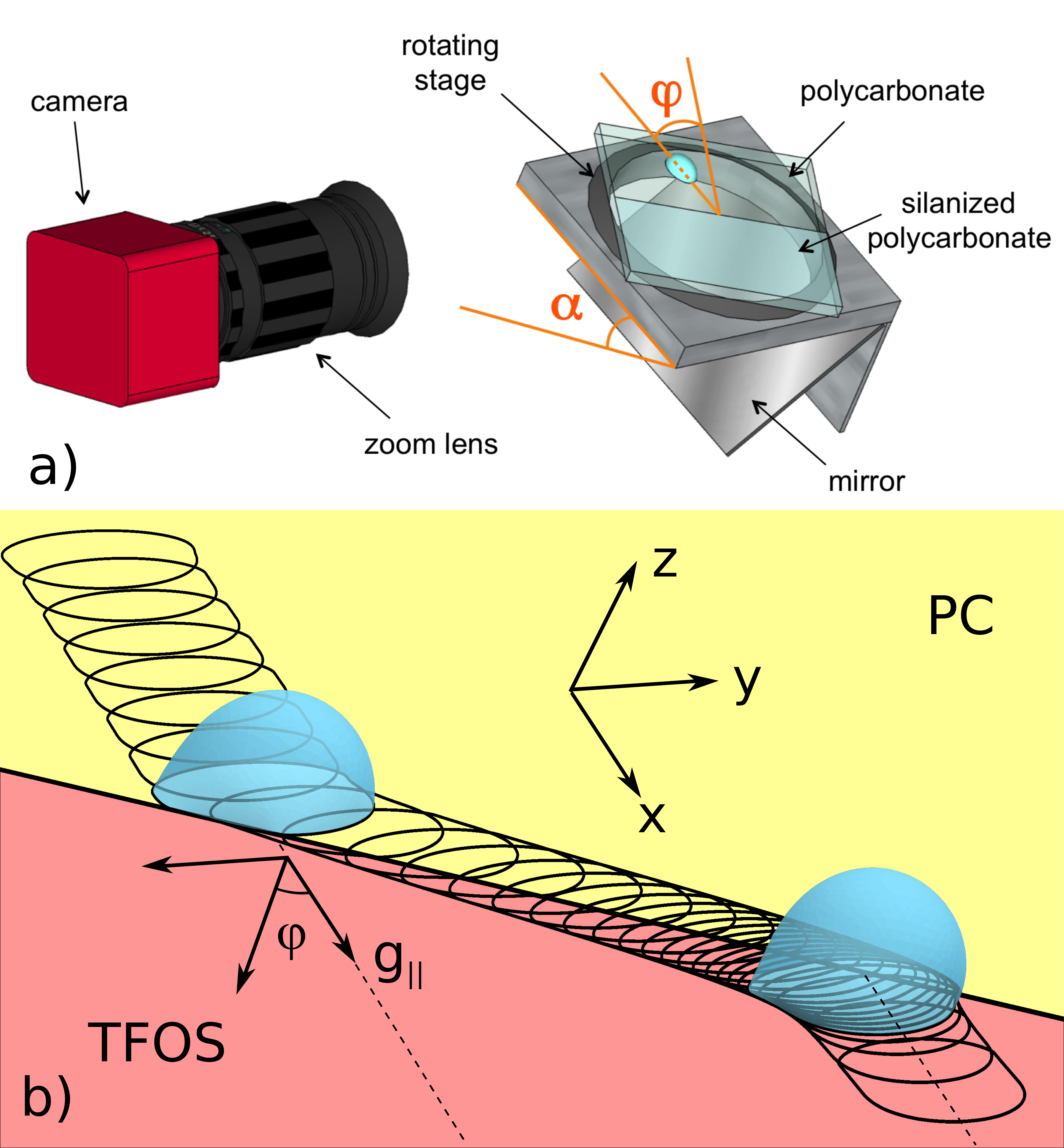}
\caption{(color online) a) Sketch of the experimental setup
indicating the inclination angle $\alpha$ with the respect to gravity 
and the tilt angle $\varphi$ with the respect to the direction of the in-plane body force.  
b) Rendering of the simulations for $\varphi=50^\circ$ and ${\sf Bo}=1.35$. 
PC (TFOS) indicates the polycarbonate (silanized) region of the wettability step.
The coordinate system $x,y,z$ in the tilted frame employed in the simulations
is indicated.
}
\label{fig1}
\end{figure}
\indent	To focus on the interplay between the
surface tension of the free interface and contact angle hysteresis, we
used drops of glycerol/water mixture, thus minimising inertial
effects \cite{tMannetje2014} and operating in a regime of over-damped
interfacial dynamics. As shown in Fig.~\ref{fig1}, we
tilted the chemical step by an arbitrary angle $\varphi$ with respect to the
direction of the in-plane body force, and mapped out a dynamic phase
diagram of the various regimes displayed by the sliding
drops. Fundamental insight into the mechanics of drop deformation was
gained by performing systematic numerical simulations based on a
contact line friction model \cite{Semprebon2014}. A typical drop
trajectory derived from these calculations is shown in
Fig.~\ref{fig1}b).

Sliding measurements were performed through the experimental setup
depicted in Fig.~\ref{fig1}a). We used drops of a 80\% w/w
glycerol/water solution with density $\varrho =1.21$
$\mathrm{g/cm}^3$, viscosity $\eta =52$ $\mathrm{cP,}$ i.e. fifty
times more viscous than pure water, and with a surface tension $\gamma
=65.3$ $\mathrm{mN/m}$ at T=$23^{\circ }C$. The substrate was a
polycarbonate (PC) slab having a thickness of $5$ mm and a side
length of $5$ cm. To realize the chemical step, half of the
original protective cover was kept as a mask for the deposition of a
molecular layer of trichloro(1H, 1H, 2H, 2H-perfluorooctyl)silane
(TFOS) from the vapor phase. The removal of the cover produced two
chemically distinct areas separated by a linear boundary. The
wettability of the two regions was characterized by measuring contact
angles of drops of the glycerol/water solution.  More precisely,
the advancing and receding contact angles on the PC region, where the
drops were initially deposited, were respectively $\theta
_{\mathrm{a,PC}}=(88\pm 2)^{\circ }$ and $\theta
_{\mathrm{r,PC}}=(63\pm 3)^{\circ }$, while those on the TFOS
covered portion, beyond the chemical step, were $\theta
_{\mathrm{a,TFOS}}=(118\pm 2)^{\circ }$ and $\theta
_{\mathrm{r,TFOS}}=(64\pm 4)^{\circ }$.

\begin{figure}[tb]
\centering \includegraphics[width=0.99\columnwidth]{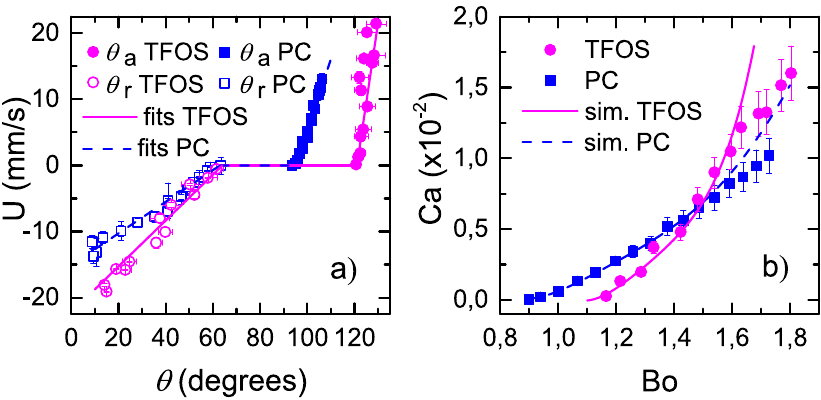}
\caption{(color online) a) Experimental measurements of the dynamic
  advancing and receding angles for PC and TFOS substrates, with
  linear fits providing the phenomenological parameters for the
  simulation. Positive (negative) velocities refer to advancing
  (receding) motion.  b) Relation between ${\sf Bo}$ and
  ${\sf Ca}$ for drops in steady motion on PC and
  TFOS. Lines are the numerical results.  }
\label{fig2}
\end{figure}

Drops of volume $V=(40\pm 2)$ $\mu\textrm{l}$ were deposited on the already
inclined plane by means of a vertically mounted syringe pump. 
The sample was placed on a manually
rotating stage with a central opening that could change the
inclination $\varphi$ of the linear chemical step. The stage was
mounted on a rotating tilting support whose inclination angle $\alpha
$ could be set by a computer with $0.1^{\circ }$ accuracy
\cite{Varagnolo2014}. The drop was lightened by two white LED
back-lights. The lateral profile of the drop was viewed with a
CMOS camera mounted along the rotation axis of
the plate and equipped with a macro zoom lens. By
moving the camera, it was possible to focus on the image of the drop
contact line reflected by a mirror mounted under the sample holder at
$45^{\circ }$ with respect to the substrate. Acquired images, where
drops appear dark on a light background, were analysed through a
custom--made LabVIEW script \cite{Toth2011}. \\
\indent Varying $\alpha $ between
$25^{\circ }$ and $60^{\circ }$, the in plane component of the body
force can be described by a Bond number ${\sf Bo}=\varrho gV^{2/3}\sin
\alpha/\gamma $ (where g is the gravity acceleration) comprised
between $0.9$ and $1.8$, as shown in Fig.~\ref{fig2}b).  The
typical velocities of the drop steadily sliding on the PC region
before touching the step ranged between $U\thicksim$ 0.1 mm/s and
$U\thicksim$ 10 mm/s, cf.~also Fig.~\ref{fig2}. Accordingly, the
maximum Weber and Capillary number of the drop with dimension
$L_0\sim V^{1/3}$ were ${\sf We}=\rho U^2 L_0/\gamma \thicksim
3\cdot10^{-3}$ and ${\sf Ca}=\eta U/\gamma\thicksim 0.01$,
respectively. Capillary waves on the drop interface that may be
excited during collision with the chemical step are quickly
damped away as the Ohnesorge number ${\sf Oh}={\sf Ca}/{\sf We}^{1/2} 
\thicksim 0.15$ is close to unity. In addition, when
crossing the chemical step the contact line velocity is further
reduced to the order of $U\thicksim 0.01$ mm/s: a regime where the
contact line mobilities on many substrates are more favourably
described by the Molecular Kinetic model\cite{Blake1997} rather than by viscous
dissipation in the fluid wedge\cite{Fetzer2010}. Hence, we
treated the drop shapes in our simulations as quasi--static and
governed by an interplay of interfacial tension, gravity, and
contact line friction. 

The assumption of a dissipation localized at the moving contact
line is not valid for interfaces in contact to substrates with low
contact angle hysteresis. In these particular cases, viscous
dissipation in the contact line region, and to a smaller extent also
in the bulk, dominate drop motion \cite{LeGrand2005}, and solutions
of the full--scale fluid dynamic problem are required
\cite{Varagnolo2013}. The essential physics of the present
experimental system, instead, can be reproduced with a minimal model
based on contact line friction that naturally combines static and
dynamic contact angle hysteresis \cite{Semprebon2014,Musterd2014}.

In the contact line friction model, the shape of the liquid interface
is assumed to be in mechanical equilibrium, and therefore at each
instant fulfils the Laplace law
\begin{equation}
  \Delta P+\varrho \,g_\parallel\,x \ =\gamma
  \kappa \label{eq:laplace}
\end{equation}
where $\Delta P$ is the pressure jump across the liquid interface at
$x=0$, $\gamma$ the liquid--vapour tension, $\kappa$ the local mean
curvature, and $g_\parallel=g\sin\alpha$ the in--plane component of
the acceleration of gravity acting in the downhill $x$--direction,
cf.~also Fig.~\ref{fig1}.  For simplicity, we neglected the component
$g_\perp=g\cos\alpha$ normal to the substrate in our calculations.
For a given contour $\Gamma$ of the contact line, being not
necessarily in mechanical equilibrium, the shape $\Sigma$ of the free
liquid interface is described by a minimum of the energy functional
\begin{equation}
  {\mathcal E}\{\Sigma \}=\gamma \,A_{\mathrm{lv}}-\varrho
  \,g_\parallel\,V\,\langle x\rangle
  \label{eq:energy}
\end{equation}%
where $A_{\mathrm{lv}}$ is the area of the free interface, while
  $\langle x \rangle$ denotes the $x$-coordinate of the drops' center
  of mass. The surface $\Sigma$ is subject to the global constraint
of a fixed liquid volume $V$. We employed the free software Surface
Evolver \cite{Brakke1996} to represent the free interface with a
triangulated mesh and minimize the energy Eq.~(\ref{eq:energy}) with
standard minimization algorithms. The motion of the contact line was
obtained by displacing the elements in contact with the substrate into
the local normal direction, according to the algorithm described in
Ref.~\cite{Semprebon2014}.

As we were operating exclusively in the limit of small contact
  line velocities in Fig.~\ref{fig2}a), we employed a linear relation
between contact line velocity and dynamic contact angles
\begin{equation}
  u=\left\{
  \begin{array}{ccc}
    u_{\mathrm{a}}(\theta -\theta _{\mathrm{a}}) & \mathrm{for} &
    \theta \geq \theta _{\mathrm{a}} \\ u_{\mathrm{r}}(\theta -\theta
    _{\mathrm{r}}) & \mathrm{for} & \theta \leq \theta _{\mathrm{r}}%
  \end{array}%
  \right. ~,  \label{eq:CL_mobility}
\end{equation}
where $u_a$ and $u_r$ were assumed to be constant over a homogeneous
portion of substrate. To match simulations with the experiments, the
phenomenological parameters $u_{\mathrm{a}}$ and $u_{\mathrm{r}}$ were
obtained by fitting the relation between contact line velocity and
dynamic contact angles derived from the experiments on PC and TFOS
substrates in Fig.~\ref{fig2}a): $u_{\mathrm{a,PC}}=61.6$ mm
s$^{-1}\mathrm{rad}^{-1}$, $u_{\mathrm{r,PC}}=13.7$ mm
s$^{-1}\mathrm{rad}^{-1}$, $u_{\mathrm{a,TFOS}}=138$ mm
s$^{-1}\mathrm{rad}^{-1}$, $u_{\mathrm{r,TFOS}}=19.9$ mm
s$^{-1}\mathrm{rad}^{-1}$.  In Fig. \ref{fig2}b) we report a
direct comparison of the relation between dimensionless drop speed $\sf Ca$ and
driving force $\sf Bo$ with the experimental data of steady drops
sliding on homogeneous PC and TFOS.  Very good agreement was found on
both substrates for ${\sf Bo} <1.6$ while numerical data
overestimate the experimental results at larger Bond numbers.  A
possible cause for the discrepancy at high Bond numbers is the
transition to a regime where the viscous dissipation in the bulk
cannot be any more neglected.

\begin{figure}[tb]
\centering \includegraphics[width=0.99\columnwidth]{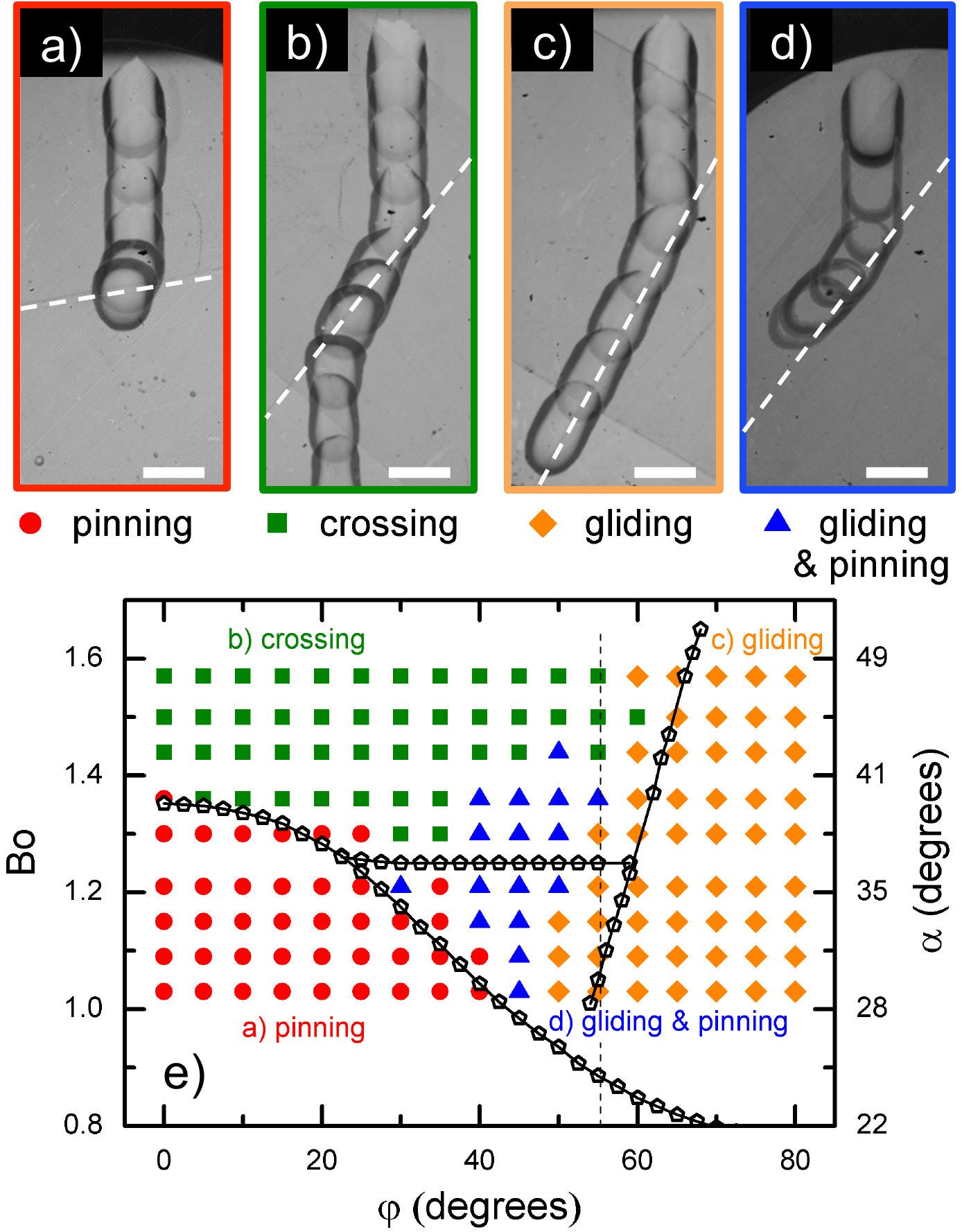}
\caption{(color online) Sequence of the four possible drop
  trajectories exhibited by a drop approaching the chemical step: a)
  the drop pins; b) the drop crosses the step; c) the drop slides
  along the step; d) the drop partially slides along the step and pins
  in a later stage.  The dashed inclined lines mark the chemical step
  and the horizontal scale bars correspond to 5 mm.  e) Dynamical
  phase diagram showing the four regions in the $\varphi$--$\sf
    Bo$ space.  Filled symbols refer to experiments, while open
  symbols are evinced from simulations. 
  Connecting lines are guide for the eye. See text for further details.}
\label{fig3}
\end{figure}

We observed four distinct scenarios for the drop approaching the
chemical step as summarized in Fig.~\ref{fig3}: a) at low $\sf
  Bo$ and $\varphi$ the drop is simply pinned at the step; b) at high
$\sf Bo$ and intermediate $\varphi$ the drop crosses the step; c)
at high $\varphi$ the drop glides along the step without crossing it;
d) at intermediate $\sf Bo$ and $\varphi$ the drop glides for a
short distance (at least 2 mm) along the step until arrest, and remains
pinned. By systematically varying the angles $\alpha$ and $\varphi$,
we constructed the dynamical phase diagram plotted in the graph of
Fig.~\ref{fig3}e). Filled symbols of different colors and shapes represent
the path followed by the drop approaching the chemical step.  No data
are reported for ${\sf Bo}\leq1$ because drops slide very slowly and
it was very difficult to distinguish the different crossing cases.
The regions identifying the four regimes are delimited by connected 
open symbols obtained from the analysis of systematic numerical simulations.
They show a good agreement with the experimental results, considering
the unavoidable presence of various sources of defects and noise in
the system.  In particular, the transition curve separating the
pinning region from the gliding and pinning one occurs at somewhat
smaller $\varphi$ values. This could be due to the presence of substrate
defects that enhance the pinning of the drop.  Furthermore, the
transition with the gliding region occurs at larger $\varphi$ values,
suggesting that most of defect are located on the TFOS side.

The results present two counter--intuitive aspects. First, one would
expect the minimum Bond number ${\sf Bo}_\mathrm{min}$ necessary
to let the drop cross the step to grow with $\varphi$, because the
component of the body force perpendicular to the step decreases as
$\cos\varphi$. Instead we observed a decrease of 
${\sf  Bo}_\mathrm{min}$ as $\varphi$ increased. In other words, the
optimal direction for crossing is not perpendicular to the step. To prove
that such unexpected result is determined by the intrinsic hysteresis,
we performed simulations in the absence of static contact angle hysteresis
by taking two different equilibrium angles at the two sides of the
step, with the drop crossing from the more hydrophilic to the more
hydrophobic.  Regardless of the chosen combination of angles, we
always observed a monotonous increase of body force proportional to
$1/\cos\varphi$, as suggested by a simple decomposition of the body
force along the component perpendicular and parallel to the straight
boundary.  In this case the dynamical phase diagram is simplified,
showing only the crossing and gliding regimes.  The second
counter--intuitive aspect is the re--entrant shape of the diagram in
the range $50^\circ\lesssim\varphi\lesssim60^\circ$: for $\varphi$
around $55^{\circ }$ (see dashed line in Fig. \ref{fig3}e)), rising 
$\sf Bo$ from low to high values we first found the transition between 
pinning and gliding. Upon increasing $\sf Bo$ we noticed a transition 
between gliding and gliding followed by pinning. Only at higher values 
of $\sf Bo$ we observed drops crossing the step. This means that 
the increase of the body force can induce pinning of a drop otherwise in motion.

\begin{figure}[tb]
\centering \includegraphics[width=0.99\columnwidth]{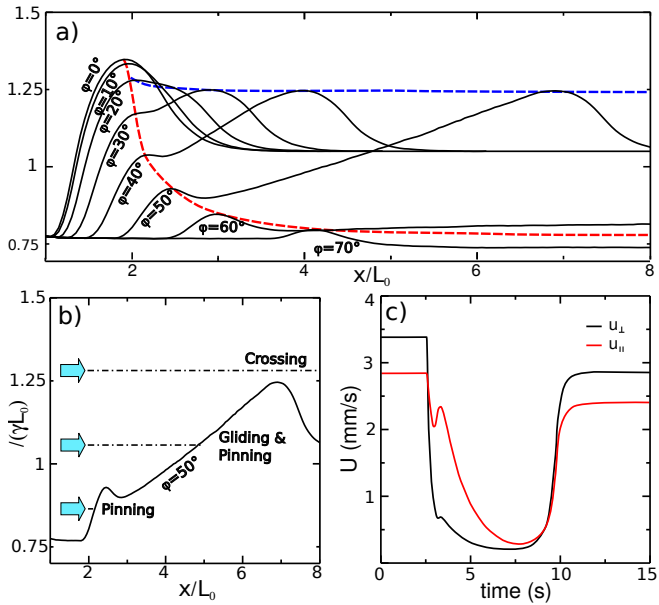}
\caption{(color online) a) Force landscape of the crossing mechanism
  for various values of $\varphi$.  The dashed lines indicate the
  trend of the two peaks present in the curves.  b) Construction of
  three crossing scenarios for the case $\varphi=50^\circ$.  c)
  Parallel and perpendicular velocity components of the drop centre of
  mass when crossing the step for $\varphi=50^\circ$ and ${\sf
      Bo}=1.35$. Series of force curves as a function of the position
  $x$ of the center of mass, normalized by $L_0=V^{1/3}$, for
  $\varphi$ varied in steps of $10^\circ$. Here the drop is initially
  placed at $x=0$, while the position of the chemical step shifted
  forward with increasing $\varphi$ for better visibility. 
  }
\label{fig4}
\end{figure} 

To gain insight into the mechanism that hinders or allows crossing of
the drop, we computed an associated force landscape. To this end we
dropped the potential energy term in Eq.~(\ref{eq:energy})
and applied instead an additional global constraint on the downhill
center of mass position, $\langle x \rangle$. The
$y$--component of the center of mass was instead allowed to
freely adapt to the value that minimizes the total energy. To
compute the force landscape we quasi--statically shifted the position
of the center of mass along the direction of the body force.  In this
case the retaining force in our simulations is given by the Lagrange
multiplier $\mu$ related to the constraint of a fixed
$\langle x \rangle$. At each increment we allowed the contact line
to relax until all the contact angles lied within the static
hysteresis interval. As shown
in the series of force curves in Fig.~\ref{fig4}a), when the drop
collides with the step, $\mu$ rises because of the contact line
deformation. For small $\varphi$ the curves exhibit a single peak, which
determines the minimum body force required to cross the step. At
larger $\varphi$ the curves show a second peak.  Here the new rise of
the force is clearly related to the shift of the drop from the low
hysteresis PC to the high hysteresis TFOS region.  Similarly to what
observed for the case of depinning from an initially circular contact
line \cite{Semprebon2014}, the second peak occurs when the drop
escapes from the self--created constriction of the contact line.  With
increasing $\varphi$ the magnitude of both peaks decreases.  While the
first peak vanishes in the limit $\varphi\rightarrow 90^\circ$, the
second peak approaches a constant value. Furthermore the distance of
the second peak from the contact point asymptotically diverges when
$\varphi$ increases, and completely disappears from our simulation
domain for $\varphi> 50^\circ$.

The analysis of the landscape allows to identify all four
experimentally observed dynamical regimes for a drop approaching
the tilted chemical step. In Fig.~\ref{fig4}b) the body force is
represented by blue arrows on the left side. If 
${\sf Bo}$ is lower than the first peak, the drop simply pins
after touching the step. If the value of ${\sf Bo}$ falls into
the interval between the two peaks, the drop glides along the step and
pins afterwards. Only if ${\sf Bo}$ exceeds the height of the
second peak, the drop crosses the step. The sequences of peaks as a
function of $\varphi$ accurately match the transition lines in
Fig.~\ref{fig3} between crossing and pinning, and between crossing and
gliding with pinning. The absence of the second peak at larger
$\varphi$ implies that the drop is entirely repelled by the step,
corresponding to the gliding observed in the experiments. The boundary
of the gliding region cannot be determined from the analysis of the
force landscape. The line displayed in Fig.~\ref{fig3} is obtained by
mapping the results of several sequences of simulations, and can be
approximately described by 
${\sf Bo}\simeq {\sf Bo}_\perp/\cos\varphi$. Here ${\sf Bo}_\perp\simeq L
(\cos\theta_{\mathrm{r}}-\cos\theta_{\mathrm{a}})$, $L$ being the
projection of the drop elongation perpendicular to the step. Such
relation implies that the body force can be split in two independent
components. While without contact angle hysteresis this is valid for 
the full range of $\varphi$, with hysteresis the decomposition is valid 
only for large $\varphi$, and in both cases it corresponds to the 
transition to gliding. Figure \ref{fig4}c), reports the two
components of the velocity profile corresponding to the trajectory
shown in Fig.~\ref{fig1}b). For drops coming from the upper
(PC) substrate, the first touch of the chemical step causes a fast
decrease of the drop velocity. After a local minimum, the velocity
slightly increases in the parallel component, due to squeezing and
alignment of the drop along the step. Consequently the drop starts a
slow side--ward shift to the lower (TFOS) substrate, which
causes a further decrease of the velocity. Eventually the drop escapes
from the chemical step and the retaining force decreases, causing a
quick acceleration, until reaching a steady motion on the TFOS
substrate. 

In summary, we have studied with experiments and numerical simulations viscous
drops crossing a tilted chemical step to address the effect of
intrinsic contact angle hysteresis and found a rich drop dynamics,
displaying four different regimes. We demonstrated that the body force
required to cross the step increases as $\sim 1/\cos\varphi$ in the
absence of intrinsic hysteresis or for sufficiently large inclinations
$\varphi$. Only without static contact angle hysteresis, the
retaining force can be split in two orthogonal components, otherwise
the coupling induced by the hysteresis gives rise to the complexity of
the observed scenario.  
In this work we have considered drops crossing from a region of lower 
to a region of higher advancing contact angles, being the receding
angle approximately the same. Correspondingly, the transition is from lower 
to higher contact angle hysteresis. The opposite scenario, involving a 
variation only in the receding angles, would not present the same 
counter-intuitive aspects, because the rise of the second peak in the force 
landscape is strictly due to the drop deformation of the front edge.
Our results show that intrinsic hysteresis
can be exploited to control drop motion in open microfluidic devices, 
particularly if they rely on breaking of the reflection symmetry.

C.S. thankfully acknowledges Halim Kusumaatmaja and Timm Kr\"uger
for inspiring discussions. M.B. and G.M. gratefully acknowledge the
Vigoni exchange program between the Ateneo Italo-Tedesco and the
German Academic Exchange Service (DAAD). We are particularly grateful
to Paolo Sartori for kind support in the set-up preparation.

\bibliographystyle{apsrev4-1} 
\bibliography{references}

\begin{thebibliography}{31}%
\makeatletter
\providecommand \@ifxundefined [1]{%
 \@ifx{#1\undefined}
}%
\providecommand \@ifnum [1]{%
 \ifnum #1\expandafter \@firstoftwo
 \else \expandafter \@secondoftwo
 \fi
}%
\providecommand \@ifx [1]{%
 \ifx #1\expandafter \@firstoftwo
 \else \expandafter \@secondoftwo
 \fi
}%
\providecommand \natexlab [1]{#1}%
\providecommand \enquote  [1]{``#1''}%
\providecommand \bibnamefont  [1]{#1}%
\providecommand \bibfnamefont [1]{#1}%
\providecommand \citenamefont [1]{#1}%
\providecommand \href@noop [0]{\@secondoftwo}%
\providecommand \href [0]{\begingroup \@sanitize@url \@href}%
\providecommand \@href[1]{\@@startlink{#1}\@@href}%
\providecommand \@@href[1]{\endgroup#1\@@endlink}%
\providecommand \@sanitize@url [0]{\catcode `\\12\catcode `\$12\catcode
  `\&12\catcode `\#12\catcode `\^12\catcode `\_12\catcode `\%12\relax}%
\providecommand \@@startlink[1]{}%
\providecommand \@@endlink[0]{}%
\providecommand \url  [0]{\begingroup\@sanitize@url \@url }%
\providecommand \@url [1]{\endgroup\@href {#1}{\urlprefix }}%
\providecommand \urlprefix  [0]{URL }%
\providecommand \Eprint [0]{\href }%
\providecommand \doibase [0]{http://dx.doi.org/}%
\providecommand \selectlanguage [0]{\@gobble}%
\providecommand \bibinfo  [0]{\@secondoftwo}%
\providecommand \bibfield  [0]{\@secondoftwo}%
\providecommand \translation [1]{[#1]}%
\providecommand \BibitemOpen [0]{}%
\providecommand \bibitemStop [0]{}%
\providecommand \bibitemNoStop [0]{.\EOS\space}%
\providecommand \EOS [0]{\spacefactor3000\relax}%
\providecommand \BibitemShut  [1]{\csname bibitem#1\endcsname}%
\let\auto@bib@innerbib\@empty
\bibitem [{\citenamefont {Bonn}\ \emph {et~al.}(2009)\citenamefont {Bonn},
  \citenamefont {Eggers}, \citenamefont {Indekeu},\ and\ \citenamefont
  {Meunier}}]{Bonn2009a}%
  \BibitemOpen
  \bibfield  {author} {\bibinfo {author} {\bibfnamefont {D.}~\bibnamefont
  {Bonn}}, \bibinfo {author} {\bibfnamefont {J.}~\bibnamefont {Eggers}},
  \bibinfo {author} {\bibfnamefont {J.}~\bibnamefont {Indekeu}}, \ and\
  \bibinfo {author} {\bibfnamefont {J.}~\bibnamefont {Meunier}},\ }\href
  {\doibase 10.1103/RevModPhys.81.739} {\bibfield  {journal} {\bibinfo
  {journal} {Rev. Mod. Phys.}\ }\textbf {\bibinfo {volume} {81}},\ \bibinfo
  {pages} {739} (\bibinfo {year} {2009})}\BibitemShut {NoStop}%
\bibitem [{\citenamefont {Qu{\'e}r{\'e}}(2008)}]{quere2008}%
  \BibitemOpen
  \bibfield  {author} {\bibinfo {author} {\bibfnamefont {D.}~\bibnamefont
  {Qu{\'e}r{\'e}}},\ }\href@noop {} {\bibfield  {journal} {\bibinfo  {journal}
  {Annu. Rev. Mater. Res.}\ }\textbf {\bibinfo {volume} {38}},\ \bibinfo
  {pages} {71} (\bibinfo {year} {2008})}\BibitemShut {NoStop}%
\bibitem [{\citenamefont {Liu}\ \emph {et~al.}(2010)\citenamefont {Liu},
  \citenamefont {Yao},\ and\ \citenamefont {Jiang}}]{liu2010}%
  \BibitemOpen
  \bibfield  {author} {\bibinfo {author} {\bibfnamefont {K.}~\bibnamefont
  {Liu}}, \bibinfo {author} {\bibfnamefont {X.}~\bibnamefont {Yao}}, \ and\
  \bibinfo {author} {\bibfnamefont {L.}~\bibnamefont {Jiang}},\ }\href@noop {}
  {\bibfield  {journal} {\bibinfo  {journal} {Chem. Soc. Rev.}\ }\textbf
  {\bibinfo {volume} {39}},\ \bibinfo {pages} {3240} (\bibinfo {year}
  {2010})}\BibitemShut {NoStop}%
\bibitem [{\citenamefont {Varagnolo}\ \emph {et~al.}(2013)\citenamefont
  {Varagnolo}, \citenamefont {Ferraro}, \citenamefont {Fantinel}, \citenamefont
  {Pierno}, \citenamefont {Mistura}, \citenamefont {Amati}, \citenamefont
  {Biferale},\ and\ \citenamefont {Sbragaglia}}]{Varagnolo2013}%
  \BibitemOpen
  \bibfield  {author} {\bibinfo {author} {\bibfnamefont {S.}~\bibnamefont
  {Varagnolo}}, \bibinfo {author} {\bibfnamefont {D.}~\bibnamefont {Ferraro}},
  \bibinfo {author} {\bibfnamefont {P.}~\bibnamefont {Fantinel}}, \bibinfo
  {author} {\bibfnamefont {M.}~\bibnamefont {Pierno}}, \bibinfo {author}
  {\bibfnamefont {G.}~\bibnamefont {Mistura}}, \bibinfo {author} {\bibfnamefont
  {G.}~\bibnamefont {Amati}}, \bibinfo {author} {\bibfnamefont
  {L.}~\bibnamefont {Biferale}}, \ and\ \bibinfo {author} {\bibfnamefont
  {M.}~\bibnamefont {Sbragaglia}},\ }\href {\doibase
  10.1103/PhysRevLett.111.066101} {\bibfield  {journal} {\bibinfo  {journal}
  {Phys. Rev. Lett.}\ }\textbf {\bibinfo {volume} {111}},\ \bibinfo {pages}
  {066101} (\bibinfo {year} {2013})}\BibitemShut {NoStop}%
\bibitem [{\citenamefont {Varagnolo}\ \emph {et~al.}(2014)\citenamefont
  {Varagnolo}, \citenamefont {Schiocchet}, \citenamefont {Ferraro},
  \citenamefont {Pierno}, \citenamefont {Mistura}, \citenamefont {Sbragaglia},
  \citenamefont {Gupta},\ and\ \citenamefont {Amati}}]{Varagnolo2014}%
  \BibitemOpen
  \bibfield  {author} {\bibinfo {author} {\bibfnamefont {S.}~\bibnamefont
  {Varagnolo}}, \bibinfo {author} {\bibfnamefont {V.}~\bibnamefont
  {Schiocchet}}, \bibinfo {author} {\bibfnamefont {D.}~\bibnamefont {Ferraro}},
  \bibinfo {author} {\bibfnamefont {M.}~\bibnamefont {Pierno}}, \bibinfo
  {author} {\bibfnamefont {G.}~\bibnamefont {Mistura}}, \bibinfo {author}
  {\bibfnamefont {M.}~\bibnamefont {Sbragaglia}}, \bibinfo {author}
  {\bibfnamefont {A.}~\bibnamefont {Gupta}}, \ and\ \bibinfo {author}
  {\bibfnamefont {G.}~\bibnamefont {Amati}},\ }\href {\doibase
  10.1021/la404502g} {\bibfield  {journal} {\bibinfo  {journal} {Langmuir}\
  }\textbf {\bibinfo {volume} {30}},\ \bibinfo {pages} {2401} (\bibinfo {year}
  {2014})}\BibitemShut {NoStop}%
\bibitem [{\citenamefont {Nakajima}\ \emph {et~al.}(2013)\citenamefont
  {Nakajima}, \citenamefont {Nakagawa}, \citenamefont {Furuta}, \citenamefont
  {Sakai}, \citenamefont {Isobe},\ and\ \citenamefont
  {Matsushita}}]{nakajima2013}%
  \BibitemOpen
  \bibfield  {author} {\bibinfo {author} {\bibfnamefont {A.}~\bibnamefont
  {Nakajima}}, \bibinfo {author} {\bibfnamefont {Y.}~\bibnamefont {Nakagawa}},
  \bibinfo {author} {\bibfnamefont {T.}~\bibnamefont {Furuta}}, \bibinfo
  {author} {\bibfnamefont {M.}~\bibnamefont {Sakai}}, \bibinfo {author}
  {\bibfnamefont {T.}~\bibnamefont {Isobe}}, \ and\ \bibinfo {author}
  {\bibfnamefont {S.}~\bibnamefont {Matsushita}},\ }\href@noop {} {\bibfield
  {journal} {\bibinfo  {journal} {Langmuir}\ }\textbf {\bibinfo {volume}
  {29}},\ \bibinfo {pages} {9269} (\bibinfo {year} {2013})}\BibitemShut
  {NoStop}%
\bibitem [{\citenamefont {Kalinin}\ \emph {et~al.}(2009)\citenamefont
  {Kalinin}, \citenamefont {Berejnov},\ and\ \citenamefont
  {Thorne}}]{Kalinin2009}%
  \BibitemOpen
  \bibfield  {author} {\bibinfo {author} {\bibfnamefont {Y.~V.}\ \bibnamefont
  {Kalinin}}, \bibinfo {author} {\bibfnamefont {V.}~\bibnamefont {Berejnov}}, \
  and\ \bibinfo {author} {\bibfnamefont {R.~E.}\ \bibnamefont {Thorne}},\
  }\href {\doibase 10.1021/la804095y} {\bibfield  {journal} {\bibinfo
  {journal} {Langmuir}\ }\textbf {\bibinfo {volume} {25}},\ \bibinfo {pages}
  {5391} (\bibinfo {year} {2009})}\BibitemShut {NoStop}%
\bibitem [{\citenamefont {Baret}\ \emph {et~al.}(2007)\citenamefont {Baret},
  \citenamefont {Decr{\'{e}}}, \citenamefont {Herminghaus},\ and\ \citenamefont
  {Seemann}}]{baret2007}%
  \BibitemOpen
  \bibfield  {author} {\bibinfo {author} {\bibfnamefont {J.~C.}\ \bibnamefont
  {Baret}}, \bibinfo {author} {\bibfnamefont {M.~M.~J.}\ \bibnamefont
  {Decr{\'{e}}}}, \bibinfo {author} {\bibfnamefont {S.}~\bibnamefont
  {Herminghaus}}, \ and\ \bibinfo {author} {\bibfnamefont {R.}~\bibnamefont
  {Seemann}},\ }\href@noop {} {\bibfield  {journal} {\bibinfo  {journal}
  {Langmuir}\ }\textbf {\bibinfo {volume} {23}},\ \bibinfo {pages} {5200}
  (\bibinfo {year} {2007})}\BibitemShut {NoStop}%
\bibitem [{\citenamefont {'t~Mannetje}\ \emph {et~al.}(2011)\citenamefont
  {'t~Mannetje}, \citenamefont {Murade}, \citenamefont {van~den Ende},\ and\
  \citenamefont {Mugele}}]{tMannetje2011}%
  \BibitemOpen
  \bibfield  {author} {\bibinfo {author} {\bibfnamefont {D.~J. C.~M.}\
  \bibnamefont {'t~Mannetje}}, \bibinfo {author} {\bibfnamefont {C.~U.}\
  \bibnamefont {Murade}}, \bibinfo {author} {\bibfnamefont {D.}~\bibnamefont
  {van~den Ende}}, \ and\ \bibinfo {author} {\bibfnamefont {F.}~\bibnamefont
  {Mugele}},\ }\href {\doibase 10.1063/1.3533362} {\bibfield  {journal}
  {\bibinfo  {journal} {Appl. Phys. Lett.}\ }\textbf {\bibinfo {volume} {98}},\
  \bibinfo {pages} {014102} (\bibinfo {year} {2011})}\BibitemShut {NoStop}%
\bibitem [{\citenamefont {'t~Mannetje}\ \emph {et~al.}(2014)\citenamefont
  {'t~Mannetje}, \citenamefont {Ghosh}, \citenamefont {Lagraauw}, \citenamefont
  {Otten}, \citenamefont {Pit}, \citenamefont {Berendsen}, \citenamefont
  {Zeegers}, \citenamefont {van~den Ende},\ and\ \citenamefont
  {Mugele}}]{tMannetje2014}%
  \BibitemOpen
  \bibfield  {author} {\bibinfo {author} {\bibfnamefont {D.~J. C.~M.}\
  \bibnamefont {'t~Mannetje}}, \bibinfo {author} {\bibfnamefont
  {S.}~\bibnamefont {Ghosh}}, \bibinfo {author} {\bibfnamefont
  {R.}~\bibnamefont {Lagraauw}}, \bibinfo {author} {\bibfnamefont
  {S.}~\bibnamefont {Otten}}, \bibinfo {author} {\bibfnamefont
  {A.}~\bibnamefont {Pit}}, \bibinfo {author} {\bibfnamefont {C.}~\bibnamefont
  {Berendsen}}, \bibinfo {author} {\bibfnamefont {J.}~\bibnamefont {Zeegers}},
  \bibinfo {author} {\bibfnamefont {D.}~\bibnamefont {van~den Ende}}, \ and\
  \bibinfo {author} {\bibfnamefont {F.}~\bibnamefont {Mugele}},\ }\href
  {\doibase 10.1038/ncomms4559} {\bibfield  {journal} {\bibinfo  {journal}
  {Nat. Commun.}\ }\textbf {\bibinfo {volume} {5}},\ \bibinfo {pages} {3559}
  (\bibinfo {year} {2014})}\BibitemShut {NoStop}%
\bibitem [{\citenamefont {Brunet}\ \emph {et~al.}(2010)\citenamefont {Brunet},
  \citenamefont {Baudoin}, \citenamefont {Matar},\ and\ \citenamefont
  {Zoueshtiagh}}]{brunet2010}%
  \BibitemOpen
  \bibfield  {author} {\bibinfo {author} {\bibfnamefont {P.}~\bibnamefont
  {Brunet}}, \bibinfo {author} {\bibfnamefont {M.}~\bibnamefont {Baudoin}},
  \bibinfo {author} {\bibfnamefont {O.~B.}\ \bibnamefont {Matar}}, \ and\
  \bibinfo {author} {\bibfnamefont {F.}~\bibnamefont {Zoueshtiagh}},\
  }\href@noop {} {\bibfield  {journal} {\bibinfo  {journal} {Physical review.
  E, Statistical, nonlinear, and soft matter physics}\ }\textbf {\bibinfo
  {volume} {81}},\ \bibinfo {pages} {036315} (\bibinfo {year}
  {2010})}\BibitemShut {NoStop}%
\bibitem [{\citenamefont {Bliznyuk}\ \emph {et~al.}(2010)\citenamefont
  {Bliznyuk}, \citenamefont {Jansen}, \citenamefont {Kooij},\ and\
  \citenamefont {Poelsema}}]{Bliznyuk2010}%
  \BibitemOpen
  \bibfield  {author} {\bibinfo {author} {\bibfnamefont {O.}~\bibnamefont
  {Bliznyuk}}, \bibinfo {author} {\bibfnamefont {H.~P.}\ \bibnamefont
  {Jansen}}, \bibinfo {author} {\bibfnamefont {E.~S.}\ \bibnamefont {Kooij}}, \
  and\ \bibinfo {author} {\bibfnamefont {B.}~\bibnamefont {Poelsema}},\ }\href
  {\doibase 10.1021/la903205e} {\bibfield  {journal} {\bibinfo  {journal}
  {Langmuir}\ }\textbf {\bibinfo {volume} {26}},\ \bibinfo {pages} {6328}
  (\bibinfo {year} {2010})}\BibitemShut {NoStop}%
\bibitem [{\citenamefont {Courbin}\ \emph {et~al.}(2007)\citenamefont
  {Courbin}, \citenamefont {Denieul}, \citenamefont {Dressaire}, \citenamefont
  {Roper}, \citenamefont {Ajdari},\ and\ \citenamefont {Stone}}]{Courbin2007}%
  \BibitemOpen
  \bibfield  {author} {\bibinfo {author} {\bibfnamefont {L.}~\bibnamefont
  {Courbin}}, \bibinfo {author} {\bibfnamefont {E.}~\bibnamefont {Denieul}},
  \bibinfo {author} {\bibfnamefont {E.}~\bibnamefont {Dressaire}}, \bibinfo
  {author} {\bibfnamefont {M.}~\bibnamefont {Roper}}, \bibinfo {author}
  {\bibfnamefont {A.}~\bibnamefont {Ajdari}}, \ and\ \bibinfo {author}
  {\bibfnamefont {H.~A.}\ \bibnamefont {Stone}},\ }\href {\doibase
  10.1038/nmat1978} {\bibfield  {journal} {\bibinfo  {journal} {Nat. Mate.}\
  }\textbf {\bibinfo {volume} {6}},\ \bibinfo {pages} {661} (\bibinfo {year}
  {2007})}\BibitemShut {NoStop}%
\bibitem [{\citenamefont {Chu}\ \emph {et~al.}(2010)\citenamefont {Chu},
  \citenamefont {Xiao},\ and\ \citenamefont {Wang}}]{Chu2010}%
  \BibitemOpen
  \bibfield  {author} {\bibinfo {author} {\bibfnamefont {K.-H.}\ \bibnamefont
  {Chu}}, \bibinfo {author} {\bibfnamefont {R.}~\bibnamefont {Xiao}}, \ and\
  \bibinfo {author} {\bibfnamefont {E.~N.}\ \bibnamefont {Wang}},\ }\href
  {\doibase 10.1038/nmat2726} {\bibfield  {journal} {\bibinfo  {journal} {Nat.
  Mater.}\ }\textbf {\bibinfo {volume} {9}},\ \bibinfo {pages} {413} (\bibinfo
  {year} {2010})}\BibitemShut {NoStop}%
\bibitem [{\citenamefont {Malvadkar}\ \emph {et~al.}(2010)\citenamefont
  {Malvadkar}, \citenamefont {Hancock}, \citenamefont {Sekeroglu},
  \citenamefont {Dressick},\ and\ \citenamefont {Demirel}}]{Malvadkar2010}%
  \BibitemOpen
  \bibfield  {author} {\bibinfo {author} {\bibfnamefont {N.~A.}\ \bibnamefont
  {Malvadkar}}, \bibinfo {author} {\bibfnamefont {M.~J.}\ \bibnamefont
  {Hancock}}, \bibinfo {author} {\bibfnamefont {K.}~\bibnamefont {Sekeroglu}},
  \bibinfo {author} {\bibfnamefont {W.~J.}\ \bibnamefont {Dressick}}, \ and\
  \bibinfo {author} {\bibfnamefont {M.~C.}\ \bibnamefont {Demirel}},\ }\href
  {\doibase 10.1038/nmat2864} {\bibfield  {journal} {\bibinfo  {journal} {Nat.
  Mater.}\ }\textbf {\bibinfo {volume} {9}},\ \bibinfo {pages} {1023} (\bibinfo
  {year} {2010})}\BibitemShut {NoStop}%
\bibitem [{\citenamefont {Semprebon}\ \emph {et~al.}(2012)\citenamefont
  {Semprebon}, \citenamefont {Herminghaus},\ and\ \citenamefont
  {Brinkmann}}]{Semprebon2012}%
  \BibitemOpen
  \bibfield  {author} {\bibinfo {author} {\bibfnamefont {C.}~\bibnamefont
  {Semprebon}}, \bibinfo {author} {\bibfnamefont {S.}~\bibnamefont
  {Herminghaus}}, \ and\ \bibinfo {author} {\bibfnamefont {M.}~\bibnamefont
  {Brinkmann}},\ }\href {\doibase 10.1039/c2sm25156f} {\bibfield  {journal}
  {\bibinfo  {journal} {Soft Matter}\ }\textbf {\bibinfo {volume} {8}},\
  \bibinfo {pages} {6301} (\bibinfo {year} {2012})}\BibitemShut {NoStop}%
\bibitem [{\citenamefont {Joanny}\ and\ \citenamefont
  {Robbins}(1990)}]{Joanny1990}%
  \BibitemOpen
  \bibfield  {author} {\bibinfo {author} {\bibfnamefont {J.~F.}\ \bibnamefont
  {Joanny}}\ and\ \bibinfo {author} {\bibfnamefont {M.~O.}\ \bibnamefont
  {Robbins}},\ }\href {\doibase 10.1063/1.458579} {\bibfield  {journal}
  {\bibinfo  {journal} {J. Chem. Phys}\ }\textbf {\bibinfo {volume} {92}},\
  \bibinfo {pages} {3206} (\bibinfo {year} {1990})}\BibitemShut {NoStop}%
\bibitem [{\citenamefont {Savva}\ and\ \citenamefont
  {Kalliadasis}(2013)}]{Savva2013}%
  \BibitemOpen
  \bibfield  {author} {\bibinfo {author} {\bibfnamefont {N.}~\bibnamefont
  {Savva}}\ and\ \bibinfo {author} {\bibfnamefont {S.}~\bibnamefont
  {Kalliadasis}},\ }\href {\doibase 10.1017/jfm.2013.201} {\bibfield  {journal}
  {\bibinfo  {journal} {J. Fluid Mech.}\ }\textbf {\bibinfo {volume} {725}},\
  \bibinfo {pages} {462} (\bibinfo {year} {2013})}\BibitemShut {NoStop}%
\bibitem [{\citenamefont {Herde}\ \emph {et~al.}(2012)\citenamefont {Herde},
  \citenamefont {Thiele}, \citenamefont {Herminghaus},\ and\ \citenamefont
  {Brinkmann}}]{Herde2012}%
  \BibitemOpen
  \bibfield  {author} {\bibinfo {author} {\bibfnamefont {D.}~\bibnamefont
  {Herde}}, \bibinfo {author} {\bibfnamefont {U.}~\bibnamefont {Thiele}},
  \bibinfo {author} {\bibfnamefont {S.}~\bibnamefont {Herminghaus}}, \ and\
  \bibinfo {author} {\bibfnamefont {M.}~\bibnamefont {Brinkmann}},\ }\href
  {\doibase 10.1209/0295-5075/100/16002} {\bibfield  {journal} {\bibinfo
  {journal} {Europhys. Lett.}\ }\textbf {\bibinfo {volume} {100}},\ \bibinfo
  {pages} {16002} (\bibinfo {year} {2012})}\BibitemShut {NoStop}%
\bibitem [{\citenamefont {Kusumaatmaja}\ and\ \citenamefont
  {Yeomans}(2007)}]{Kusumaatmaja2007}%
  \BibitemOpen
  \bibfield  {author} {\bibinfo {author} {\bibfnamefont {H.}~\bibnamefont
  {Kusumaatmaja}}\ and\ \bibinfo {author} {\bibfnamefont {J.~M.}\ \bibnamefont
  {Yeomans}},\ }\href {\doibase 10.1021/la062082w} {\bibfield  {journal}
  {\bibinfo  {journal} {Langmuir}\ }\textbf {\bibinfo {volume} {23}},\ \bibinfo
  {pages} {956} (\bibinfo {year} {2007})}\BibitemShut {NoStop}%
\bibitem [{\citenamefont {Sbragaglia}\ \emph {et~al.}(2014)\citenamefont
  {Sbragaglia}, \citenamefont {Biferale}, \citenamefont {Amati}, \citenamefont
  {Varagnolo}, \citenamefont {Ferraro}, \citenamefont {Mistura},\ and\
  \citenamefont {Pierno}}]{Sbragaglia2013}%
  \BibitemOpen
  \bibfield  {author} {\bibinfo {author} {\bibfnamefont {M.}~\bibnamefont
  {Sbragaglia}}, \bibinfo {author} {\bibfnamefont {L.}~\bibnamefont
  {Biferale}}, \bibinfo {author} {\bibfnamefont {G.}~\bibnamefont {Amati}},
  \bibinfo {author} {\bibfnamefont {S.}~\bibnamefont {Varagnolo}}, \bibinfo
  {author} {\bibfnamefont {D.}~\bibnamefont {Ferraro}}, \bibinfo {author}
  {\bibfnamefont {G.}~\bibnamefont {Mistura}}, \ and\ \bibinfo {author}
  {\bibfnamefont {M.}~\bibnamefont {Pierno}},\ }\href {\doibase
  10.1103/PhysRevE.89.012406} {\bibfield  {journal} {\bibinfo  {journal} {Phys.
  Rev. E}\ }\textbf {\bibinfo {volume} {89}},\ \bibinfo {pages} {1} (\bibinfo
  {year} {2014})}\BibitemShut {NoStop}%
\bibitem [{\citenamefont {Cavalli}\ \emph {et~al.}(2015)\citenamefont
  {Cavalli}, \citenamefont {Musterd},\ and\ \citenamefont
  {Mugele}}]{Cavalli2015}%
  \BibitemOpen
  \bibfield  {author} {\bibinfo {author} {\bibfnamefont {A.}~\bibnamefont
  {Cavalli}}, \bibinfo {author} {\bibfnamefont {M.}~\bibnamefont {Musterd}}, \
  and\ \bibinfo {author} {\bibfnamefont {F.}~\bibnamefont {Mugele}},\ }\href
  {\doibase 10.1103/PhysRevE.91.023013} {\bibfield  {journal} {\bibinfo
  {journal} {Phys. Rev. E}\ }\textbf {\bibinfo {volume} {91}},\ \bibinfo
  {pages} {1} (\bibinfo {year} {2015})}\BibitemShut {NoStop}%
\bibitem [{\citenamefont {Nilsson}\ and\ \citenamefont
  {Rothstein}(2012)}]{Nilsson2012}%
  \BibitemOpen
  \bibfield  {author} {\bibinfo {author} {\bibfnamefont {M.~A.}\ \bibnamefont
  {Nilsson}}\ and\ \bibinfo {author} {\bibfnamefont {J.~P.}\ \bibnamefont
  {Rothstein}},\ }\href {\doibase 10.1063/1.4723866} {\bibfield  {journal}
  {\bibinfo  {journal} {Phys. Fluids}\ }\textbf {\bibinfo {volume} {24}},\
  \bibinfo {pages} {062001} (\bibinfo {year} {2012})}\BibitemShut {NoStop}%
\bibitem [{\citenamefont {Suzuki}\ \emph {et~al.}(2008)\citenamefont {Suzuki},
  \citenamefont {Nakajima}, \citenamefont {Tanaka}, \citenamefont {Sakai},
  \citenamefont {Hashimoto}, \citenamefont {Yoshida}, \citenamefont
  {Kameshima},\ and\ \citenamefont {Okada}}]{Suzuki2008}%
  \BibitemOpen
  \bibfield  {author} {\bibinfo {author} {\bibfnamefont {S.}~\bibnamefont
  {Suzuki}}, \bibinfo {author} {\bibfnamefont {A.}~\bibnamefont {Nakajima}},
  \bibinfo {author} {\bibfnamefont {K.}~\bibnamefont {Tanaka}}, \bibinfo
  {author} {\bibfnamefont {M.}~\bibnamefont {Sakai}}, \bibinfo {author}
  {\bibfnamefont {A.}~\bibnamefont {Hashimoto}}, \bibinfo {author}
  {\bibfnamefont {N.}~\bibnamefont {Yoshida}}, \bibinfo {author} {\bibfnamefont
  {Y.}~\bibnamefont {Kameshima}}, \ and\ \bibinfo {author} {\bibfnamefont
  {K.}~\bibnamefont {Okada}},\ }\href {\doibase 10.1016/j.apsusc.2007.07.171}
  {\bibfield  {journal} {\bibinfo  {journal} {Appl. Surf. Science}\ }\textbf
  {\bibinfo {volume} {254}},\ \bibinfo {pages} {1797} (\bibinfo {year}
  {2008})}\BibitemShut {NoStop}%
\bibitem [{\citenamefont {Semprebon}\ and\ \citenamefont
  {Brinkmann}(2014)}]{Semprebon2014}%
  \BibitemOpen
  \bibfield  {author} {\bibinfo {author} {\bibfnamefont {C.}~\bibnamefont
  {Semprebon}}\ and\ \bibinfo {author} {\bibfnamefont {M.}~\bibnamefont
  {Brinkmann}},\ }\href {\doibase 10.1039/c3sm51959g} {\bibfield  {journal}
  {\bibinfo  {journal} {Soft Matter}\ }\textbf {\bibinfo {volume} {10}},\
  \bibinfo {pages} {3325} (\bibinfo {year} {2014})}\BibitemShut {NoStop}%
\bibitem [{\citenamefont {Tóth}\ \emph {et~al.}(2011)\citenamefont {Tóth},
  \citenamefont {Ferraro}, \citenamefont {Chiarello}, \citenamefont {Pierno},
  \citenamefont {Mistura}, \citenamefont {Bissacco},\ and\ \citenamefont
  {Semprebon}}]{Toth2011}%
  \BibitemOpen
  \bibfield  {author} {\bibinfo {author} {\bibfnamefont {T.}~\bibnamefont
  {Tóth}}, \bibinfo {author} {\bibfnamefont {D.}~\bibnamefont {Ferraro}},
  \bibinfo {author} {\bibfnamefont {E.}~\bibnamefont {Chiarello}}, \bibinfo
  {author} {\bibfnamefont {M.}~\bibnamefont {Pierno}}, \bibinfo {author}
  {\bibfnamefont {G.}~\bibnamefont {Mistura}}, \bibinfo {author} {\bibfnamefont
  {G.}~\bibnamefont {Bissacco}}, \ and\ \bibinfo {author} {\bibfnamefont
  {C.}~\bibnamefont {Semprebon}},\ }\href {\doibase 10.1021/la2001249}
  {\bibfield  {journal} {\bibinfo  {journal} {Langmuir}\ }\textbf {\bibinfo
  {volume} {27}},\ \bibinfo {pages} {4742} (\bibinfo {year}
  {2011})}\BibitemShut {NoStop}%
\bibitem [{\citenamefont {Blake}\ \emph {et~al.}(1997)\citenamefont {Blake},
  \citenamefont {Clarke}, \citenamefont {DeConinck},\ and\ \citenamefont
  {DeRuijter}}]{Blake1997}%
  \BibitemOpen
  \bibfield  {author} {\bibinfo {author} {\bibfnamefont {T.~D.}\ \bibnamefont
  {Blake}}, \bibinfo {author} {\bibfnamefont {a.}~\bibnamefont {Clarke}},
  \bibinfo {author} {\bibfnamefont {J.}~\bibnamefont {DeConinck}}, \ and\
  \bibinfo {author} {\bibfnamefont {M.~J.}\ \bibnamefont {DeRuijter}},\
  }\href@noop {} {\bibfield  {journal} {\bibinfo  {journal} {Langmuir}\
  }\textbf {\bibinfo {volume} {13}},\ \bibinfo {pages} {2164} (\bibinfo {year}
  {1997})}\BibitemShut {NoStop}%
\bibitem [{\citenamefont {Fetzer}\ and\ \citenamefont
  {Ralston}(2010)}]{Fetzer2010}%
  \BibitemOpen
  \bibfield  {author} {\bibinfo {author} {\bibfnamefont {R.}~\bibnamefont
  {Fetzer}}\ and\ \bibinfo {author} {\bibfnamefont {J.}~\bibnamefont
  {Ralston}},\ }\href@noop {} {\bibfield  {journal} {\bibinfo  {journal} {J.
  Phys. Chem. C}\ }\textbf {\bibinfo {volume} {114}},\ \bibinfo {pages} {12675}
  (\bibinfo {year} {2010})}\BibitemShut {NoStop}%
\bibitem [{\citenamefont {{Le Grand}}\ \emph {et~al.}(2005)\citenamefont {{Le
  Grand}}, \citenamefont {Daerr},\ and\ \citenamefont {Limat}}]{LeGrand2005}%
  \BibitemOpen
  \bibfield  {author} {\bibinfo {author} {\bibfnamefont {N.}~\bibnamefont {{Le
  Grand}}}, \bibinfo {author} {\bibfnamefont {A.}~\bibnamefont {Daerr}}, \ and\
  \bibinfo {author} {\bibfnamefont {L.}~\bibnamefont {Limat}},\ }\href
  {\doibase 10.1017/S0022112005006105} {\bibfield  {journal} {\bibinfo
  {journal} {J. Fluid Mech.}\ }\textbf {\bibinfo {volume} {541}},\ \bibinfo
  {pages} {293} (\bibinfo {year} {2005})}\BibitemShut {NoStop}%
\bibitem [{\citenamefont {Musterd}\ \emph {et~al.}(2014)\citenamefont
  {Musterd}, \citenamefont {van Steijn}, \citenamefont {Kleijn},\ and\
  \citenamefont {Kreutzer}}]{Musterd2014}%
  \BibitemOpen
  \bibfield  {author} {\bibinfo {author} {\bibfnamefont {M.}~\bibnamefont
  {Musterd}}, \bibinfo {author} {\bibfnamefont {V.}~\bibnamefont {van Steijn}},
  \bibinfo {author} {\bibfnamefont {C.~R.}\ \bibnamefont {Kleijn}}, \ and\
  \bibinfo {author} {\bibfnamefont {M.~T.}\ \bibnamefont {Kreutzer}},\ }\href
  {\doibase 10.1103/PhysRevLett.113.066104} {\bibfield  {journal} {\bibinfo
  {journal} {Phys. Rev. Lett.}\ }\textbf {\bibinfo {volume} {113}},\ \bibinfo
  {pages} {066104} (\bibinfo {year} {2014})}\BibitemShut {NoStop}%
\bibitem [{\citenamefont {Brakke}(1996)}]{Brakke1996}%
  \BibitemOpen
  \bibfield  {author} {\bibinfo {author} {\bibfnamefont {K.~A.}\ \bibnamefont
  {Brakke}},\ }\href {\doibase 10.1098/rsta.1996.0095} {\bibfield  {journal}
  {\bibinfo  {journal} {Philos. T. Roy. Soc. A}\ }\textbf {\bibinfo {volume}
  {354}},\ \bibinfo {pages} {2143} (\bibinfo {year} {1996})}\BibitemShut
  {NoStop}%
\end{thebibliography}%

\end{document}